# Accurate Reporting of Ion Time-of-Flight during HiPIMS with Gated Front-End Mass Spectrometry

Nathan Rodkey[1,*], Jyotish Patidar[1], Kerstin Thorwarth[1], Sebastian Siol[1,*]

[1]*Laboratory for Surface Science and Coating Technologies, Empa – Swiss Federal Laboratories for Materials Science and Technology, Switzerland*

Corresponding authors: nathan.rodkey@empa.ch, sebastian.siol@empa.ch

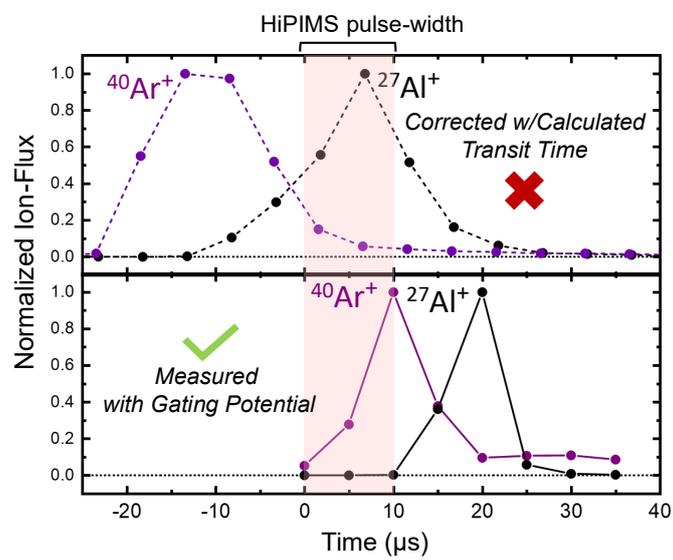

**Abstract**: The quality of high-power impulse magnetron sputtering (HiPIMS) deposited films can often improve through the effective use of metal-ion acceleration, requiring precise measurements of time-of-flight (ToF). These measurements are commonly done using time- and energy-resolved mass spectrometry but require careful consideration of the transit time of ions inside. The transit time is typically calculated by considering the travel length in various parts of the spectrometer (e.g. from orifice to detector), but errors associated with these estimations can lead to nonphysical values in a HiPIMS process (e.g. negative ToFs). Here we report a practical approach to determine ion ToF experimentally, using a bipolar HiPIMS power supply to synchronize a gating pulse to the front-end of a HIDEN Analytical EQP-300 mass spectrometer, placed at the working distance. The ToF is measured by applying a +70 V bias to repel ions, and a 5 μs gating pulse of 0 V to accept them. To prevent interference with the HiPIMS plasma, a grounded shield is placed in front of the mass-spec head with a variable slit-opening (0.5–3 mm). The effectiveness of the shielding is verified by Langmuir probe measurements, noting negligible shifts in plasma potential for a DC sputter discharge. The gate is then synchronized to a HiPIMS pulse and data collected at 5 μs intervals by adjusting the pulse delay. Measurements of the time-of-flights of $Ar^+$, $Al^+$, $Sc^+$, $Y^+$, and $W^+$ ions are presented; $Al^+$ and $Ar^+$ ions were also compared to ToF calculated using mass spectrometry flight tube equations.

## INTRODUCTION

In high-power impulse magnetron sputtering (HiPIMS), short, μs-scale voltage pulses induce high peak currents on a sputter target. The resulting increase in plasma-density facilitates the ionization of sputtered species resulting in high ionized flux fractions (up to 90%) when compared to a standard direct current (DC) process (~3-4%).[1] The sputtered ions arrive at different times on the substrate, with process gas ions (e.g. Ar) often arriving first.[2–4] This difference in time-of-flight (ToF) can be used to selectively accelerate species, applying a substrate bias only when specific ions arrive. This procedure is termed metal-ion synchronization (MIS). MIS-HiPIMS is an effective method for controlling stress, solubility, and defects (e.g. Ar incorporation), and has been shown to produce films of higher quality when compared to conventional ion acceleration using DC potentials.[5–15] However, effective metal-ion synchronization requires precise measurements of the ToF of ions, a requirement that is exacerbated when using short voltage pulses (e.g. 10 μs) and/or when accelerating metal ions that are light and have thus short ToFs (e.g. $Al^+$).

ToF measurements are challenging and require dedicated plasma characterization equipment. Retarding Field Energy Analyzers (RFEA) have been shown for time-resolved measurements but are not ideal for ToF measurements as they cannot distinguish between ionic species and/or process gasses.[16,17] Instead, the most common way to measure ToF is with time- and energy- resolved mass spectrometry.[18–27] However, a challenge associated with these measurements comes from the calculation of transit time of ions inside the mass spectrometer, needed to accurately report the ToF.

During a synchronized HiPIMS experiment, the value of interest is the time-resolved ion-flux in the substrate position (i.e. at the working distance) relative to the start of the HiPIMS pulse at $t = 0$. Since the ion flux is measured at a detector on the far end of the mass spectrometer, the measured ToF, $t_{measured}$ consists of the actual ToF of the ions from the target to the working distance $t_{actual}$ and the transit time through the mass-spectrometer, $t_{transit}$:

$$t_{\text{measured}} = t_{\text{actual}} + t_{\text{transit}} \qquad Eq.\ 1$$

The transit time can be estimated using the travel length in varying parts of the spectrometer, as well as the interaction of ions with varying electrostatic optics at a constant kinetic energy. Shown in *Eq. 2* is an example equation provided by *HIDEN Analytical* for calculating the transit time in an *EQP-300* mass spectrometer. It is important to note, that the transit time inside the mass spectrometer can be several times larger than the actual ToF from the target to the working distance. For example, $Ar^+$ ions reported in this work had a measured ToF of 10 μs and a transit time of 113 μs. Therefore, even small errors in the transit time can lead to significant errors in the reported ToF of ions. This, in turn, can lead to nonphysical values of $t_{actual}$ in a HiPIMS process, such as negative ToFs, or metal ions arriving to the substrate before process gas ions. Consequently, groups that report ToF often clarify that the measurements are estimates or treat ToF data qualitatively.[18–25] While this approach helps in identifying trends, it does not give ToF data precise enough for establishing efficient MIS processes, adding extra process work to determine the best conditions for applying an accelerating bias. As a result, transit times are best determined experimentally.

This can be done by using a gated voltage pulse applied to the driven front-end (or orifice) of the mass-spectrometer. Here, a positive voltage repels the ions most of the time, whereas a smaller gating pulse (at negative or zero potential to ground) allows ions to pass at a well-defined time interval. This concept was first demonstrated by the group of *J.* Bradley at the University of Manchester, attaching a double-gridded mesh for the gating of ions.[28] At the time, this concept was used to measure ion-energy density functions (IEDFs) and later on time-resolved mass spectrometry characterization of pulsed plasmas, previously inaccessible as the transit time of ions within the mass-spectrometer limited the measurable frequency of pulses to <10 kHz.[29,30] By applying electrostatic gating potentials to conductive meshes in front of the orifice, time-resolved IEDFs were determined without the need to reference the inherent time delay caused by the transit time of ions through the flight tube.

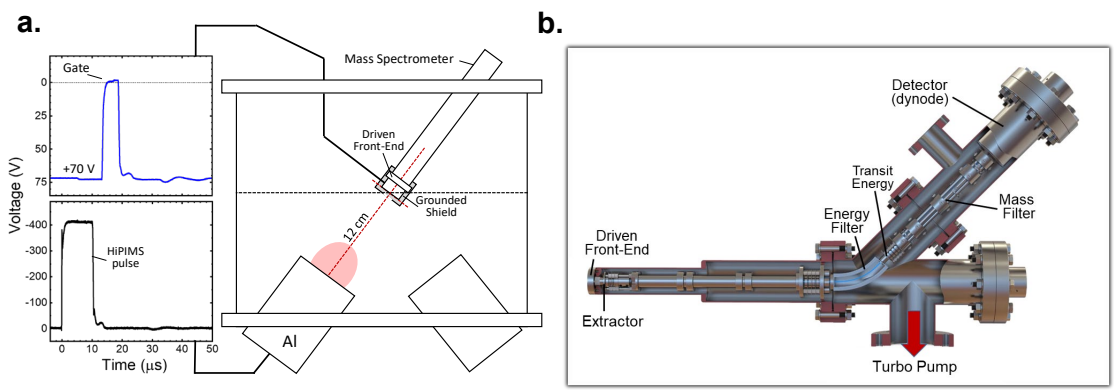

**Figure 1. a.** Schematic of a sputter system and electrical signals used in a gated ToF measurement. A *HIDEN Analytical EQP-300* mass spectrometer is used with a driven front-end where a synchronized 5 μs gating pulse is applied. An example of the gating pulse and a 10 μs HiPIMS discharge used in this work are shown on the left. To prevent interference with the plasma, a grounded shield is placed over the driven front-end. **b.** Schematic of the EQP-300 mass spectrometer, with the labelled relevant optics used in *Eq. 2* for the transit time in tube equation. Source: *HIDEN Analytical*.[31]

While their solution provides accurate results with high-temporal resolution, it is not available off the shelf and the necessary customizations to the mass spectrometer are rather involved. Here we report on a practical solution that can be used on standard mass spectrometers with a driven front-end option. First, the orifice of a mass-spec must be moved to the desired substrate position, after which a positive bias is maintained to screen ions, and a 0 V gating pulse to accept them. This is shown schematically in **Figure 1a**. In this way, only ions that have reached the desired substrate-target distance during the gating pulse are measured, and no correction for transit tube time-of-flight is needed. In this example, we show how a *HIDEN EQP-300* mass spectrometer, a common mass-spec model used for measuring ToF in many published works,[18–25] can be modified by attaching a grounded shield to its driven front-end. This is done to prevent interference with the plasma, as a positive bias will substantially increase the plasma potential, changing the plasma dynamics and potentially making the discharge unstable. This technique was demonstrated by our group for several studies of AlN-based materials, and the 3D drawings for this grounded shield are provided in the *Supporting Information*.[6,7,13,15] We note that for groups that don't have a driven front-end equipped to their mass-spectrometers, the work shown here demonstrates that a mesh assembly is not necessary and that similar gating can be achieved using a metal plate with a small orifice (~50 μm). The ToFs of $Ar^+$, $Al^+$, $Sc^+$, Y+, and $W^+$ ions were measured in this way and, for $Al^+$ and $Ar^+$ ions, compared to those calculated using mass spectrometry flight tube equations.

**EXPERIMENTAL DETAILS**

The experiments were carried out in a commercial sputter chamber (AJA ATC-1800) in confocal sputter up geometry. A HiPIMS discharge was then ignited on an unbalanced magnetron using a 2" target. The magnetron was in an open-field magnetic configuration with the surrounding magnetrons. The magnetron was pointed directly at the orifice of the mass spectrometer for the measurements. The HiPIMS deposition was carried out using a pulse-width of 10 μs, a frequency of 2500 Hz, and a working pressure of 5 μbar using an Ionautics HiPSTER 1 Bipolar power supply. The discharge was power controlled, and peak-current densities ($J_{pk}$) were regulated by increasing the power, with the power-densities used for each material detailed in **Table S1**. The process gas (Ar 6.0) was routed directly to the chimney of the magnetron. These parameters are kept constant for all materials studied in this work.

The ToF measurements were performed by recording ion energy distribution functions in a *HIDEN Analytical EQP-300* mass spectrometer (50 μm orifice) placed at a 12 cm working distance directly aligned with the deposition axis of the sputter gun. A bipolar HiPIMS power supply (Ionautics HiPSTER 1 Bipolar) was used to provide the synchronized gating pulse at the front-end of the mass spectrometer. ToF was then measured by applying a +70 V bias to repel ions and a 5 μs gating pulse at 0 V to accept them. The gate was synchronized to the HiPIMS discharge by feeding a trigger signal to a synchronization unit, and data was collected at 5 μs intervals by adjusting the time delay of this pulse, tracked on an oscilloscope. This is shown schematically in **Figure 1**. The ion-energy distribution function (IEDF) is measured in time-averaged mode for each gate, and integrated to determine a relative, total ion count. To prevent interference of the driven front-end (kept at +70 V) with the HiPIMS plasma, a grounded shield is placed in front of the mass-spec head with a variable slit-opening between 0.5 – 3 mm, kept electrically isolated with a 1 mm PEEK (polyether ether ketone) spacer. The 3D .step files of the grounded shield and PEEK spacer used to modify a *HIDEN EQP-300* are included in the *Supporting Information*, with pictures of the shield components and 3D models shown in **Figure S1, S2, S3**. The effectiveness at screening ions with a +70 V bias is shown in **Figure S4** where the IDF of $Ar^+$ and $Al^+$ ions are shown at different screening biases. Note, that a bump in intensity at the applied bias is observed, affecting more strongly the process gas ions ($Ar^+$); this is likely related to an increase in plasma potential between the grounded shield and driven front-end, where the energy of thermalized $Ar^+$ ions is increased to match the plasma potential. As a result of this bump in intensity, the ToFs of metal ions in this work are measured by integrating IEDFs up to 50 eV, while Ar ToFs are integrated up to 30 eV to avoid some of this background signal. Langmuir probe measurements were performed using a 5 mm long, 380 μm diameter Pt probe, connected to an *Impedans* control unit.

## RESULTS AND DISCUSSION

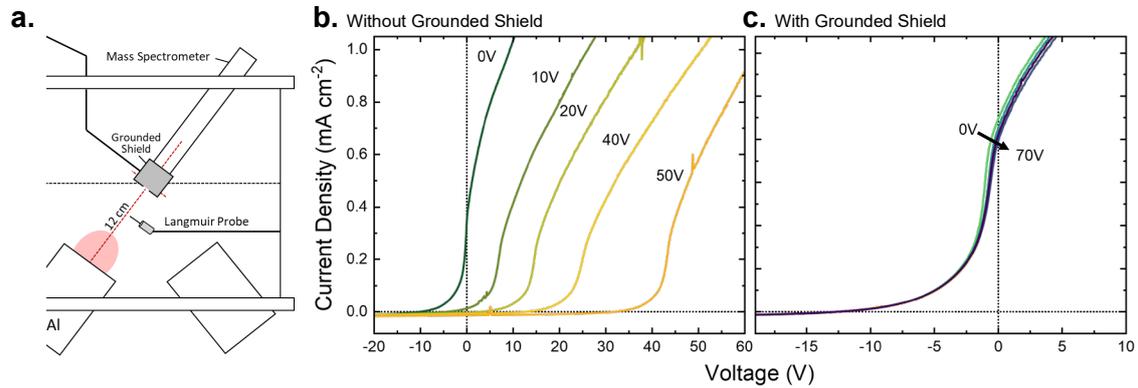

**Figure 2.** A Langmuir probe was placed between the mass-spectrometer and an Al sputter target, shown schematically in panel **a**. Then, a Langmuir scan was swept without a grounded shield (panel **b**) and with a grounded shield (panel **c**). Without the grounded shield, the plasma potential rises with the applied bias, before becoming unstable past +50 V. With a grounded shield the plasma potential shift is negligible and the HiPIMS discharge remains stable.

Applying a positive voltage on the mass-spectrometer front-end interferes with the discharge, by raising the plasma potential in the vicinity of the magnetron. This is seen in Langmuir probe scans in **Figure 2**. The position of the Langmuir probe is shown schematically in **Figure 2a**. When a positive voltage is applied to the front-end of the mass-spectrometer, the plasma potential rises, indicated by a shift of the JV scan towards the right (**Fig. 2b**). These measurements were performed using a DC plasma, as the plasma dynamics of a HiPIMS plasma can be quite complex and the discharge itself could not be maintained when applying a front-end bias larger than +20 V. When a grounded shield is used, the front-end can be driven up to +70 V with negligible differences in the plasma potential **(Fig. 2c)**.

With the grounded shield installed, ToF measurements were collected with and without a gating voltage applied to the front-end of the mass spectrometer. When measuring ToF without a front-end gate, but in a time-resolved manner, gating of the signal typically happens at the detector such that the arrival time of ions must be corrected for their transit time and interactions with varying electrostatic optics, the most relevant of which are shown in **Figure 1b**. This way of determining the ion ToF is prevalent in literature.[18–25] *Eq. (2)* shows the transit tube time equation provided by *HIDEN* for correcting ToF in an *EQP-300*.

$$d_{ext}\frac{\sqrt{2m}}{\sqrt{e(K_{ion}-V_{ext})}+\sqrt{|eV_{endcap}|}} + d_{en}\sqrt{\frac{m}{|2e(V_{endcap}-V_{cylinder}-K_{ion})|}} + d_{mass}\sqrt{\frac{m}{|2eV_{te}|}} + d_{det}\sqrt{\frac{2m}{|eV_{dyn}|}} \quad Eq.\ 2$$

Where d denotes the length of various elements ($d_{en}$ = energy filter, $d_{ext}$ = length of extractor, $d_{mass}$ = mass filter, $d_{det}$ = detector), m the ion mass, $K_{ion}$ the initial ion energy, and V the voltage of different elements ($V_{ext}$ = extractor, $V_{dyn}$ = dynode, $V_{te}$ = transit energy, $V_{endcap}$ = driven front-end, $V_{cylinder}$ = main cylinder).

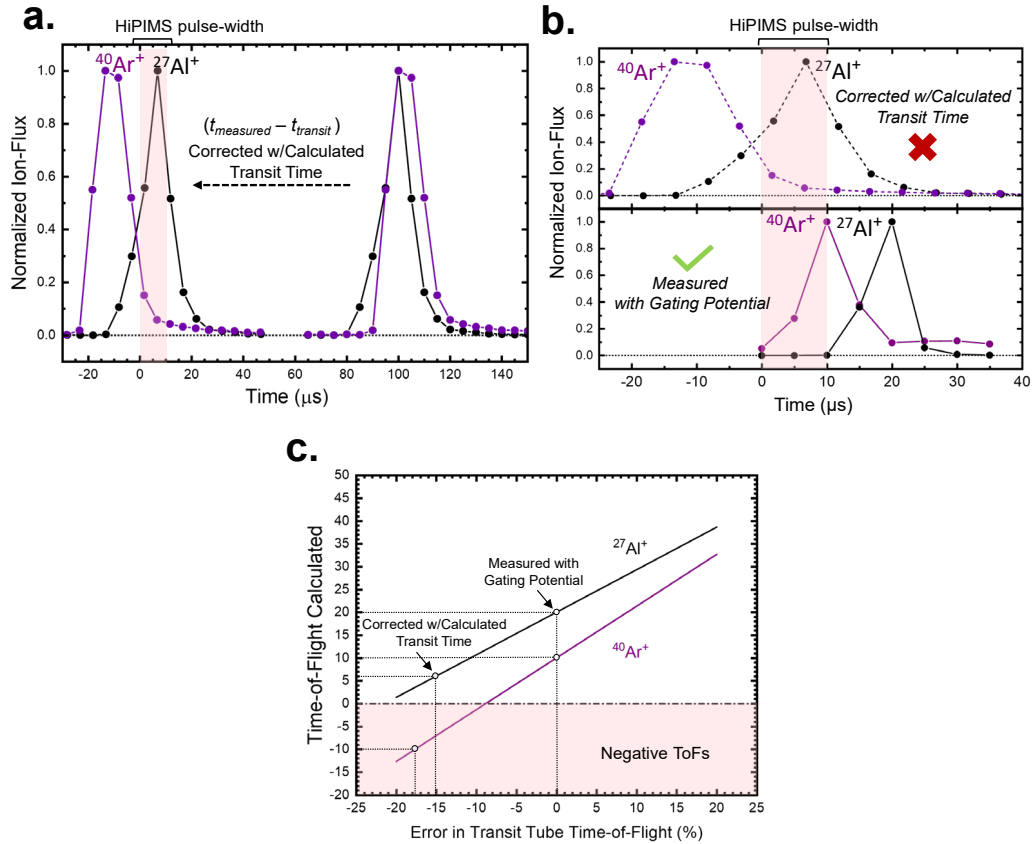

**Figure 3.** In panel **a.** the time-of-flight (ToF) of Ar⁺ and Al⁺ ions measured using a *HIDEN Analytical EQP-300* is displayed. Shown are the measured and actual ToF, $t_{measured}$ and $t_{actual}$, respectively. $t_{actual}$ is corrected using the calculated transit time from the tube equation (*Eq. 2*) leading to negative ToF values for both, Ar⁺ and Al⁺. **b.** Comparison of the gated front-end and transit time corrected approaches. Notably, the corrected ToF is negative and the approximate delay between the arrival of Ar⁺ and Al⁺ ions is distended when compared to the gated front-end approach (16 vs. 10 μs). This is also seen in panel **c.** where an estimation of the impact of error in the transit time in tube equation is shown on the corrected ToFs. Marked are ToFs measured using both the gated and corrected for transit time approaches.

Unfortunately, the error margin when applying this correction is significant and can lead to nonphysical values. This is especially relevant when considering short voltage pulses and light metal ions (which have a short ToF). This is shown in **Figure 3**, for an Al HiPIMS process, employing a 10 μs pulse-width. ToF measurements performed without a front-end gate and subsequently corrected with the transit time in tube equation (*Eq. 1*) are shown in **Figure 3a**. After the correction, negative time of flights are calculated for the process gas ions, with an onset of Al⁺ ions measured before the start of the HiPIMS discharge at *t* = 0. In **Figure 3b** a comparison of ToF calculated using the transit time equation (top panel) and measured using a gated front-end (bottom panel) are shown. Additional to the nonphysical values calculated via the transit time equation, the difference between Ar⁺ and Al⁺ ToF is distended when compared to the gated front-end approach (~16 vs. 10 μs). For the HiPIMS discharge shown here, accurate metal-ion synchronization is not possible when simply correcting time of flights with the transit time equation.

The emphasize the impact error in the transit tube ToF calculations can have, **Figure 3c** shows the calculated ToF of Al$^+$ and Ar$^+$ ions with different transit-times (referenced as a % error to the ToF calculated using *Eq. 2*). This shows that errors as small as 5% can introduce 5 μs shifts in estimated ToFs. Significant considering the measured 10 and 20 μs ToFs of Ar$^+$ and Al$^+$ respectively. Additionally, the deviation of the corrected ToFs from the measured, gated ToFs suggest that a component of error depends on the mass of the ionic species. Seen by the estimated -18% vs. -15 % errors for respective Ar$^+$ and Al$^+$ ions.

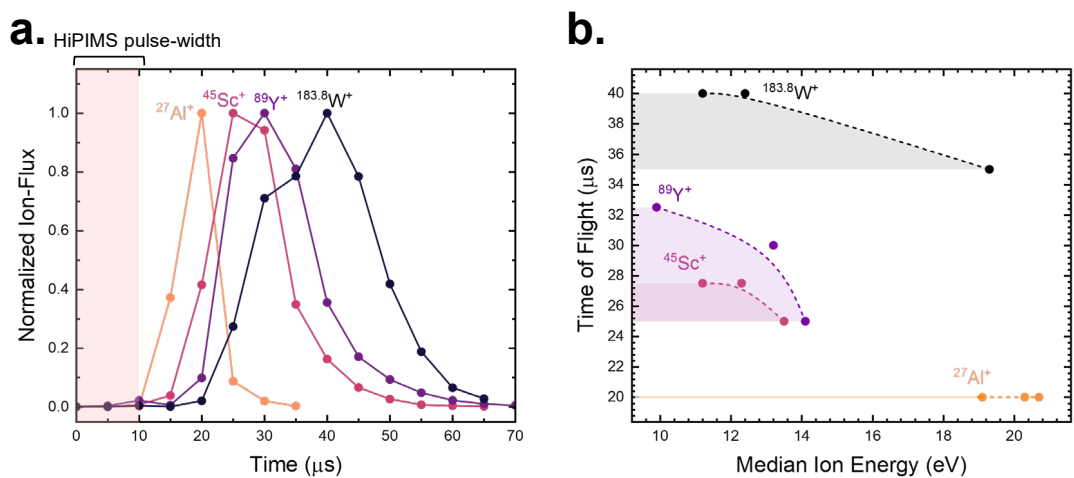

**Figure 4.** In **a.** the ToF measured for Al$^+$, Sc$^+$, Y$^+$, and W$^+$ ions with an $J_{pk}$ density of 0.59 Acm$^{-2}$ are shown for a working distance of 12 cm. In **b.** the ToF peak position is plotted as a function of median ion energy, with higher ion energies leading to shorter ToFs. The shaded areas give an idea for what range of ToFs each of these materials tend to have.

To extend the study, ToF of Al$^+$, Sc$^+$, Y$^+$, and W$^+$ ions were measured at peak current densities of 0.15, 0.59, and 0.89 Acm$^{-2}$. An IDEF is collected at each gating voltage and integrated to obtain the relative intensity of ions at that time-step. The gating bias has a 5 μs pulse-width, and the center of this pulse is defined as the time-step. The normalized ion-flux intensities for a peak-current density of 0.59 Acm$^{-2}$ are shown in **Figure 4a**, showing a clear dependence of the ToF on atomic mass (amu). In **Figure 4b**, the ToF is plotted vs. median ion energy, showing how increased ion energy reduces ToF. This trend was also extrapolated for amu vs. peak-current, shown in **Figure S6**. For most species, higher current densities — typically leading to higher ion energies — trend towards lower ToFs. However, for W (amu 183.8), this trend seems to invert, with longer ToFs observed at higher current-densities. This was correlated to a decrease in median ion energy at higher $J_{pk}$ current densities, possibly caused by an increase in collisions in the plasma, lowering the overall energy of ions.[14] As higher current-densities are accessed by increasing power to the sputter target (frequency is fixed), this problem may be exacerbated by the increasing plasma densities. Similar trends with amu were also shown by *Greczynski et al.* using much higher pulse-widths (150 μs) and a *HIDEN EQP1000*.[18] While they were corrected using the same transit time in tube equation shown above, negative ToFs were not seen, likely due to the larger pulse-widths used in their work.

## CONCLUSION

Accurate measurements of time-of-flight (ToF) are needed for the selective acceleration of metal-ions, particularly relevant in metal-ion synchronized HiPIMS (MIS-HiPIMS). As these processes gain popularity, more groups face the challenge of accurate ToF measurements, needed for proper synchronization. However, not much literature is available on best practices for the accurate determination of ToF. At present, the most common way to measure ToF is by correcting the transit time of ions in the mass-spectrometer, influenced by various tube lengths and interactions with electrostatic optics. These corrections can be inaccurate for the timescales needed in MIS-HiPIMS, particularly for short voltage pulses (e.g. 10 μs) and light ions where ToFs are short (e.g. $Al^+$). Instead, we suggest ToF measurements to be made by first positioning the mass-spectrometer at the substrate position and then applying a gating voltage to accept ions at precise time intervals. This involves a bipolar HiPIMS power supply and a straightforward modification, applicable to any EQP-series mass-spectrometer with a driven front-end. In this example, this is done by modifying a *HIDEN EQP-300* mass-spectrometer, equipping it with a grounded shield to prevent interference with the plasma. The 3D .step files for these pieces are included in the *Supporting Information*. We also note that if no driven front-end is present, the provided 3D files can be modified to include an orifice plate (50 μm) and PEEK spacer where a gating pulse could then be applied, an example of this is shown in **Figure S7.** The procedure here is easily applicable to existing setups in laboratories around the world and should facilitate more robust and quantitative ToF measurements in the future.

## ACKNOWLEDEGMENTS

The authors would like to acknowledge Sasa Vranjkovic and the Empa workshop for help with the design and construction of the custom parts. Holger Kersten is acknowledged for helpful discussion. Funding by the SNSF projects 10004403 and 196980 is gratefully acknowledged.

## AUTHOR DECLARATIONS

**Conflict of Interest**

The authors have no conflicts to disclose.

**Data Availability**

Data is available at reasonable request from the corresponding authors. 3D files pertaining to the design and construction of the grounded shield presented in this work can be found in the *Supporting Information*.

# References

(1) Sarakinos, K.; Alami, J.; Konstantinidis, S. High Power Pulsed Magnetron Sputtering: A Review on Scientific and Engineering State of the Art. *Surface and Coatings Technology*. February 25, 2010, pp 1661–1684. https://doi.org/10.1016/j.surfcoat.2009.11.013.

(2) Anders, A. Tutorial: Reactive High Power Impulse Magnetron Sputtering (R-HiPIMS). *J Appl Phys* **2017**, *121* (17), 171101. https://doi.org/10.1063/1.4978350.

(3) Greczynski, G.; Petrov, I.; Greene, J. E.; Hultman, L. Paradigm Shift in Thin-Film Growth by Magnetron Sputtering: From Gas-Ion to Metal-Ion Irradiation of the Growing Film. *Journal of Vacuum Science & Technology A* **2019**, *37* (6), 060801. https://doi.org/10.1116/1.5121226.

(4) Macák, K.; Kouznetsov, V.; Schneider, J.; Helmersson, U.; Petrov, I. Ionized Sputter Deposition Using an Extremely High Plasma Density Pulsed Magnetron Discharge. *Journal of Vacuum Science & Technology A: Vacuum, Surfaces, and Films* **2000**, *18* (4), 1533–1537. https://doi.org/10.1116/1.582380.

(5) Greczynski, G.; Lu, J.; Johansson, M. P.; Jensen, J.; Petrov, I.; Greene, J. E.; Hultman, L. Role of Ti N+ and Al N+ Ion Irradiation (N=1, 2) during Ti 1-XAl XN Alloy Film Growth in a Hybrid HIPIMS/Magnetron Mode. *Surf Coat Technol* **2012**, *206* (19–20), 4202–4211. https://doi.org/10.1016/j.surfcoat.2012.04.024.

(6) Patidar, J.; Sharma, A.; Zhuk, S.; Lorenzin, G.; Cancellieri, C.; Sarott, M. F.; Trassin, M.; Thorwarth, K.; Michler, J.; Siol, S. Improving the Crystallinity and Texture of Oblique-Angle-Deposited AlN Thin Films Using Reactive Synchronized HiPIMS. *Surf Coat Technol* **2023**, *468*. https://doi.org/10.1016/j.surfcoat.2023.129719.

(7) Patidar, J.; Thorwarth, K.; Schmitz-Kempen, T.; Kessels, R.; Siol, S. Deposition of Highly-Crystalline AlScN Thin Films Using Synchronized HiPIMS-from Combinatorial Screening to Piezoelectric Devices. *Phys. Review Materials* **2024**, *8* (9). https://doi.org/tps://doi.org/10.1103/PhysRevMaterials.8.095001.

(8) Greczynski, G.; Lu, J.; Johansson, M.; Jensen, J.; Petrov, I.; Greene, J. E.; Hultman, L. Selection of Metal Ion Irradiation for Controlling Ti 1-XAl XN Alloy Growth via Hybrid HIPIMS/Magnetron Co-Sputtering. *Vacuum* **2012**, *86* (8), 1036–1040. https://doi.org/10.1016/j.vacuum.2011.10.027.

(9) Greczynski, G.; Mráz, S.; Hultman, L.; Schneider, J. M. Selectable Phase Formation in VAlN Thin Films by Controlling Al+ Subplantation Depth. *Sci Rep* **2017**, *7* (1). https://doi.org/10.1038/s41598-017-17846-5.

(10) Greczynski, G.; Lu, J.; Jensen, J.; Bolz, S.; Kölker, W.; Schiffers, C.; Lemmer, O.; Greene, J. E.; Hultman, L. A Review of Metal-Ion-Flux-Controlled Growth of Metastable TiAlN by HIPIMS/DCMS Co-Sputtering. *Surf Coat Technol* **2014**, *257*, 15–25. https://doi.org/10.1016/j.surfcoat.2014.01.055.

(11) Tang, J. F.; Huang, S. Y.; Chen, I. H.; Shen, G. L.; Chang, C. L. Effects of Synchronous Bias Mode and Duty Cycle on Microstructure and Mechanical Properties of AlTiN Coatings Deposited via HiPIMS. *Coatings* **2023**, *13* (9). https://doi.org/10.3390/coatings13091512.

(12) Gui, B.; Hu, H.; Zhou, H.; Zhang, T.; Liu, X.; Ma, Z.; Xian, C. Influence of Synchronized Pulse Bias on the Microstructure and Properties of CrSiN Nano-Composite Ceramic Films Deposited by MIS-HiPIMS. *Ceram Int* **2024**, *50* (17), 31576–31588. https://doi.org/10.1016/j.ceramint.2024.05.465.

(13) Messi, F.; Patidar, J.; Rodkey, N.; Dräyer, C. W.; Trassin, M.; Siol, S. Ferroelectric AlScN Thin Films with Enhanced Polarization and Low Leakage Enabled by High-Power Impulse Magnetron Sputtering. *APL Mater* **2025**, *13* (5). https://doi.org/10.1063/5.0267904.

(14) Shimizu, T.; Takahashi, K.; Boyd, R.; Viloan, R. P.; Keraudy, J.; Lundin, D.; Yang, M.; Helmersson, U. Low Temperature Growth of Stress-Free Single Phase α-W Films Using HiPIMS with Synchronized Pulsed Substrate Bias. *J Appl Phys* **2021**, *129* (15). https://doi.org/10.1063/5.0042608.

(15) Patidar, J.; Pshyk, O.; Thorwarth, K.; Sommerhäuser, L.; Siol, S. Low Temperature Deposition of Functional Thin Films on Insulating Substrates Enabled by Selective Ion Acceleration Using Synchronized Floating Potential HiPIMS. *Nature Communications* **2025**, *16* (1). https://doi.org/10.1038/s41467-025-59911-y.

(16) Chen, Z.; Blakeney, J.; Carruth, M.; Ventzek, P. L. G.; Ranjan, A. Time-Resolved Ion Energy Distribution in Pulsed Inductively Coupled Argon Plasma with/without DC Bias. *Journal of Vacuum Science & Technology B* **2022**, *40* (3). https://doi.org/10.1116/6.0001737.

(17) Gahan, D.; Dolinaj, B.; Hayden, C.; Hopkins, M. B. Retarding Field Analyzer for Ion Energy Distribution Measurement through a Radio-Frequency or Pulsed Biased Sheath. In *Plasma Processes and Polymers*; 2009; Vol. 6. https://doi.org/10.1002/ppap.200931607.

(18) Greczynski, G.; Mráz, S.; Schneider, J. M.; Hultman, L. Metal-Ion Subplantation: A Game Changer for Controlling Nanostructure and Phase Formation during Film Growth by Physical Vapor Deposition. *J Appl Phys* **2020**, *127* (18). https://doi.org/10.1063/1.5141342.

(19) El Farsy, A.; Boivin, D.; Noel, C.; Hugon, R.; Cuynet, S.; Bougdira, J.; De Poucques, L. Ionized Particle Transport in Reactive HiPIMS Discharge: Correlation between the Energy Distribution Functions of Neutral and Ionized Atoms. *Plasma Sources Sci Technol* **2021**, *30* (6). https://doi.org/10.1088/1361-6595/ac02b4.

(20) Kagerer, S.; Zauner, L.; Wojcik, T.; Kolozsvári, S.; Kozák, T.; Čapek, J.; Zeman, P.; Riedl, H.; Mayrhofer, P. H. Reactive HiPIMS Deposition of Al-Oxide Thin Films Using W-Alloyed Al Targets. *Surf Coat Technol* **2021**, *422* (February). https://doi.org/10.1016/j.surfcoat.2021.127467.

(21) Zauner, L.; Bahr, A.; Kozák, T.; Čapek, J.; Wojcik, T.; Hunold, O.; Kolozsvári, S.; Zeman, P.; Mayrhofer, P. H.; Riedl, H. Time-Averaged and Time-Resolved Ion Fluxes Related to Reactive HiPIMS Deposition of Ti-Al-N Films. *Surf Coat Technol* **2021**, *424* (August), 25–33. https://doi.org/10.1016/j.surfcoat.2021.127638.

(22) Sanekata, M.; Nakagomi, Y.; Hirayama, M.; Nishida, H.; Nishimiya, N.; Tona, M.; Yamamoto, H.; Tsukamoto, K.; Fuke, K.; Ohshimo, K.; Koyasu, K.; Misaizu, F. Time-of-Flight Mass Spectrometry Diagnostics in Deep Oscillation Magnetron Sputtering (DOMS) of Titanium. *J Appl Phys* **2022**, *131* (24). https://doi.org/10.1063/5.0089592.

(23) Du, H.; Zanáška, M.; Helmersson, U.; Lundin, D. On Selective Ion Acceleration in Bipolar HiPIMS: A Case Study of (Al,Cr)2O3 Film Growth. *Surf Coat Technol* **2023**, *454* (December 2022). https://doi.org/10.1016/j.surfcoat.2022.129153.

(24) Oks, E.; Anders, A. Lawrence Berkeley National Laboratory (Lbnl). *J Appl Phys* **2008**, 185–186.

(25) Bohlmark, J.; Lattemann, M.; Gudmundsson, J. T.; Ehiasarian, A. P.; Aranda Gonzalvo, Y.; Brenning, N.; Helmersson, U. The Ion Energy Distributions and Ion Flux Composition from a High Power Impulse Magnetron Sputtering Discharge. *Thin Solid Films* **2006**, *515* (4), 1522–1526. https://doi.org/10.1016/j.tsf.2006.04.051.

(26) Yamashita, Y.; Niwayama, M. Principles and Instrumentation. *Application of Near Infrared Spectroscopy in Biomedicine* **2013**, *30* (September), 1–19. https://doi.org/10.1007/978-1-4614-6252-1_1.

(27) Pareige, C.; Lefebvre-Ulrikson, W.; Vurpillot, F.; Sauvage, X. Time-of-Flight Mass Spectrometry and Composition Measurements. In *Atom Probe Tomography: Put Theory Into Practice*; Elsevier Inc., 2016; pp 123–154. https://doi.org/10.1016/B978-0-12-804647-0.00005-X.

(28) Karkari, S. K.; Bäcker, H.; Forder, D.; Bradley, J. W. A Technique for Obtaining Time- and Energy-Resolved Mass Spectroscopic Measurements on Pulsed Plasmas. *Meas Sci Technol* **2002**, *13* (9), 1431–1436. https://doi.org/10.1088/0957-0233/13/9/308.

(29) Voronin, S. A.; Alexander, M. R.; Bradley, J. W. Time-Resolved Measurements of the Ion Energy Distribution Function in a Pulsed Discharge Using a Double Gating Technique. *Meas Sci Technol* **2005**, *16* (12), 2446–2452. https://doi.org/10.1088/0957-0233/16/12/007.

(30) Voronin, S. A.; Clarke, G. C. B.; Čada, M.; Kelly, P. J.; Bradley, J. W. An Improved Method for IEDF Determination in Pulsed Plasmas and Its Application to the Pulsed Dc Magnetron. *Meas Sci Technol* **2007**, *18* (7), 1872–1876. https://doi.org/10.1088/0957-0233/18/7/012.

(31) HiDEN Analytical. *EQP Series*. https://www.hidenanalytical.com/products/thin-films-plasma-and-surface-engineering/eqp-series/.

# Supporting Information: Accurate Reporting of Ion Time-of-Flight during HiPIMS with Gated Front-End Mass Spectrometry

Nathan Rodkey[1,*], Jyotish Patidar[1], Kerstin Thorwarth[1], Sebastian Siol[1,*]

[1]*Laboratory for Surface Science and Coating Technologies, Empa – Swiss Federal Laboratories for Materials Science and Technology, Switzerland*

Corresponding authors: nathan.rodkey@empa.ch, sebastian.siol@empa.ch

*Table S1* The tabulated process parameters used for measuring ion-energy density functions (IEDFs) at different peak-current densities. These different peak-current densities were accessed by increasing power to the sputter target.

| Material | Power (Wcm$^{-2}$) | $J_{pk}$ (Acm$^{-2}$) | Frequency (Hz) | Pulse-Width (µs) | Pressure (µbar) |
|---|---|---|---|---|---|
| Al | 1.63 | 0.15 | 2500 | 10 | 5 |
| Al | 5.43 | 0.59 | 2500 | 10 | 5 |
| Al | 8.88 | 0.89 | 2500 | 10 | 5 |
| Sc | 0.35 | 0.15 | 2500 | 10 | 5 |
| Sc | 2.47 | 0.59 | 2500 | 10 | 5 |
| Sc | 4.20 | 0.89 | 2500 | 10 | 5 |
| Y | 0.49 | 0.15 | 2500 | 10 | 5 |
| Y | 3.46 | 0.59 | 2500 | 10 | 5 |
| Y | 5.68 | 0.89 | 2500 | 10 | 5 |
| W | 1.78 | 0.15 | 2500 | 10 | 5 |
| W | 6.42 | 0.59 | 2500 | 10 | 5 |
| W | 9.87 | 0.89 | 2500 | 10 | 5 |

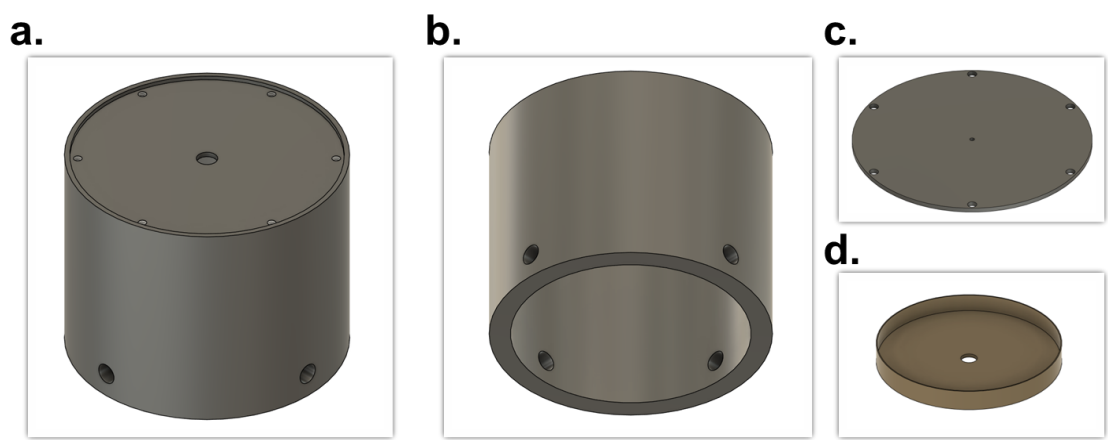

**Figure S1.** 3D model of the grounded shield components: the main body **a.** front-tilt **b.** back-tilt, **c.** the orifice plate which screws onto the main body, and **c.** the PEEK spacer that fits between the main body and the mass-spec head.

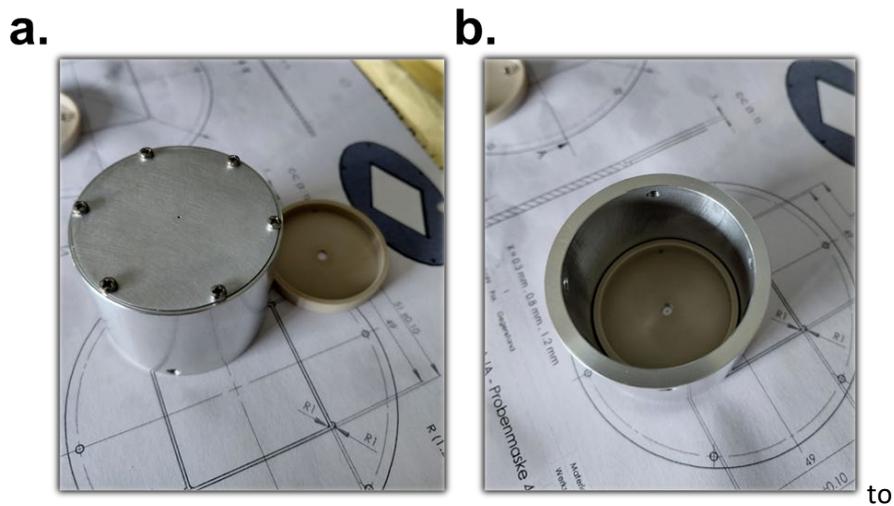

**Figure S2.** The assembled grounded shiled components, with the main body and orifice-plate shown in panel **a.** and the PEEK spacer placed inside the grounded shiled in panel **c.**

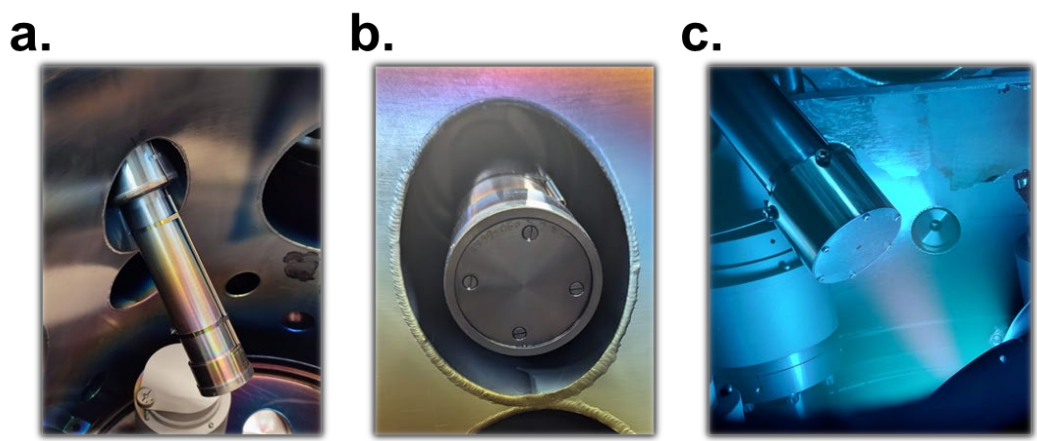

**Figure S3.** Photos of the HIDEN EQP-300 mass-spec with driven front-end **a.** and **b.** without the grounded shield assembly and **c.** with the grounded shield assembly.

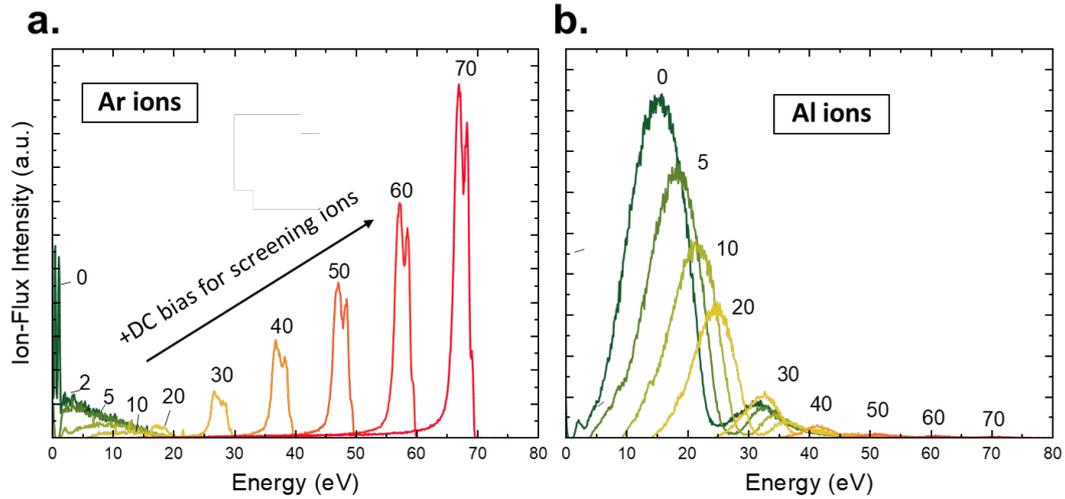

**Figure S4.** A positive, constant DC bias (0 – 70 V) is applied to the front-end of the mass-spectrometer, with a grounded shield, and the corresponding ion energy distribution functions (IEDF) are measured for **a.** Ar and **b.** Al ions. Notably, a bump at the applied +DC bias is observed, significantly more pronounced in the case of Ar ions and expected to be a result of the increased plasma potential between the grounded shield and the driven front-end of the mass-spec, which can accelerate ambient thermalized ions.

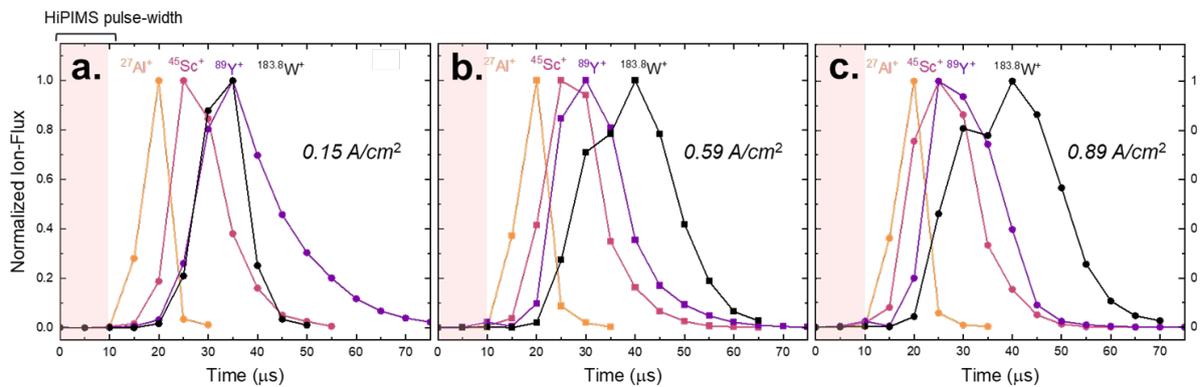

**Figure S5.** The time-of-flight for ions in a HiPIMS discharge with different peak-current densities, following similar trends in atomic mass as shown in **Figure 4**. Shaded in red is the length of the HiPIMS voltage pulse on the target.

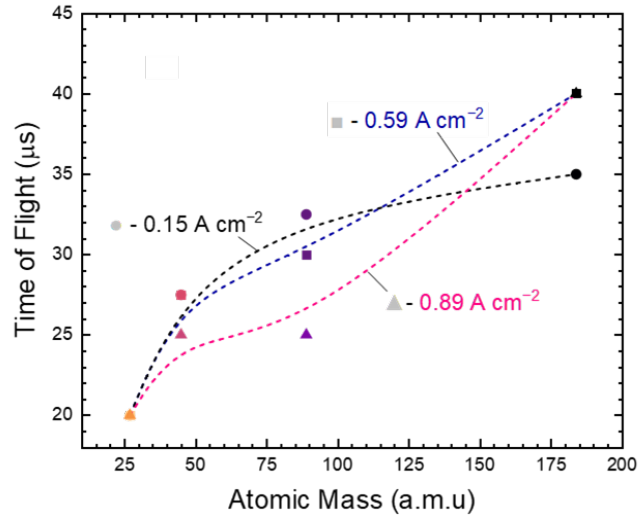

**Figure S6.** The time-of-flight (ToF) for three peak-current densities is shown, with a clear dependence on atomic mass (amu). Generally, an increase in peak-current density ($J_{pk}$) is accompanied by an increase in ion energy, increasing the speed of ions and reducing the ToF. However, for W (amu = 183.8), this trend seems to inverse. **Figure 4** shows that the relationship of ToF to ion energy is preserved. Consequently, we attribute this inversion to an incresae in plasma density at higher $J_{pk}$ currents, leading to increased scatter events and thus reduced ion kinetic energies.

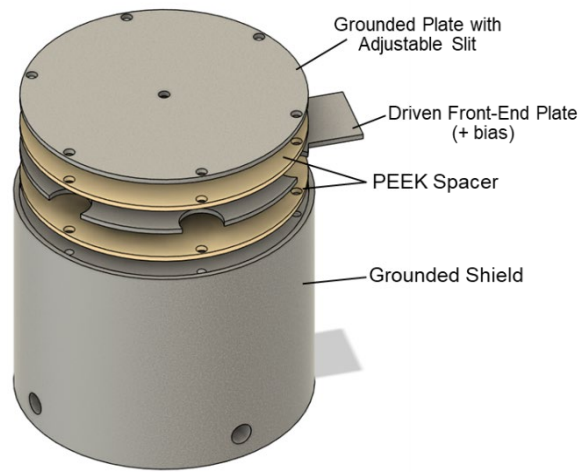

**Figure S7.** Above, an example of modifying the grounded shield presented in this work to also act as a driven front-end; useful in cases where mass spectrometers don't have an option to apply a bias to the front. Here, we show that the 3D files (provided in the *Supporting Information*) of the grounded shield assembly can be modified, placing a driven front-end plate sandwhiched between two PEEK spacers to maintain electrical isolation. Careful consideration should be taken to avoid shorting this piece to the rest of the assembly. A tab on the driven front-end plate provides easy access for electrical connections. The recommended orifice size for this positive plate would be between 0.5 – 1 mm.